\documentclass[12pt]{iopart}

\usepackage[T1]{fontenc}

\usepackage[latin1]{inputenc}

\usepackage{graphicx}

\usepackage{multirow}

\usepackage[english]{babel}

\usepackage{amssymb}

\usepackage{amsfonts}

\usepackage{subfig}

\usepackage{epsfig}

\begin{document}


\def\be{\begin{equation}}
\def\ee{\end{equation}}
\def\bea{\begin{eqnarray}}
\def\eea{\end{eqnarray}}
\def\tr{{\rm tr}\, }
\def\nn{\nonumber \\}
\def\e{{\rm e}}
\title[Cosmological evolution in viable modified gravity]{Cosmological evolution, future singularities, {\it Little Rip} and {\it Pseudo-Rip} in viable f(R) theories and their scalar-tensor counterpart}

\author{Diego S\'aez-G\'omez\footnote{E-mail: diego.saez@ehu.es}}%

\address{Fisika Teorikoaren eta Zientziaren Historia Saila, Zientzia eta Teknologia Fakultatea, Euskal Herriko Unibertsitatea, 644 Posta Kutxatila, 48080 Bilbao, Spain, EU}
%


\begin{abstract}
Modified $f(R)$ gravity is one of the most promising candidates for dark energy, and even for the unification of the whole cosmological evolution, including the inflationary phase. Within this class of theories, the so-called viable modified gravities represent  realistic  theories that are capable of reproducing late-time acceleration, and satisfy strong constraints at local scales, where General Relativity is recovered. The present manuscript deals with the analysis of the cosmological evolution for some of these models, which indicates that the evolution may enter into a phantom phase, but the behavior may be asymptotically stable. Furthermore, the scalar-tensor equivalence of $f(R)$ gravity is considered, which provides useful information about the possibility of the occurrence of a future singularity.  The so-called {\it Little Rip} and {\it Pseudo-Rip} are also studied in the framework of this class of modified gravities. 
\end{abstract}

\maketitle

\section{Introduction}

Since the discovery by two independent groups of the accelerated expansion of the universe through observations of the luminosity distance of Supernovae IA, and additionally supported by some other independent observations (such as the Cosmic Microwave anisotropies), a great effort has been made to understand what kind of mechanism is producing this anomalous behavior on the expansion of the universe, encompassing the mystery under the name of dark energy. Within this framework, a lot of candidates have been suggested, either a new field such as quintessence models or vector fields, or some modifications of General Relativity (GR) which would show effects only at cosmological scales (for a review on modified gravities see \cite{0601213}). \\

Over recent years the study of modified gravity theories has become very popular, and a major point for part of the scientific community, in order to understand the problem and the possibilities to reconstruct a gravitational theory capable of reproducing the dark energy epoch, and even the inflationary phase. In this sense, $f(R)$ theories seem to be the best candidate because of  their simplicity, by generalizing the Hilbert-Einstein action to a generic function of the Ricci scalar, and ease of reconstructing a particular action that reproduces the correct cosmological evolution (see Refs.~\cite{Ref5,Goheer:2009ss,gr-qc/0607118,0908.1269,MyF(R)1}).  Some of these models  reproduce exactly the same behavior as $\Lambda$CDM model (see Ref.~\cite{gr-qc/0607118}), or are capable of mimicking  a cosmological constant at the current epoch and to unify the entire cosmological history by reproducing  also inflation (see \cite{0908.1269,f(R)deSitter}). Furthermore,  as suggested by observations, the possibility that the dark energy behaves as a phantom fluid, i.e.  with an equation of state (EoS) parameter less than -1 currently or in the near future, is not excluded and has been widely explored (see Ref.~\cite{phantom,Sahni:2002dx}). In such a case, the universe evolution may enter in a phantom phase, producing a super-accelerating expansion that may end in a future singularity, as the so-called {\it Big Rip} (for a classification of future singularities, see Ref.~\cite{Nojiri:2005sx}, and more recently Ref.~\cite{Dabrowski:2009kg}). The cosmological singularities have been widely explored, as they content important information on the structure of a particular spacetime and its topology (see Ref.~\cite{barrow}). In this sense, some $f(R)$ models cross the phantom barrier and  lead to future singularities \cite{Nojiri:2008fk}. \\

Moreover, the high precision of local gravity tests imposes strong constraints on any modification of GR (see Ref.~\cite{f(R)GRtests}).  Nevertheless, some viable $f(R)$ models have been proposed, which are capable of reproducing a realistic cosmological evolution, and prevent violations of local gravitational tests and instabilities in the presence of matter distributions (see Refs.~\cite{f(R)viable1,f(R)viable2,f(R)viableOthers}), and even to reproduce the inflationary phase (see Ref.~\cite{Saez-Gomez1}). Nevertheless, viable $f(R)$ gravity may enter into a phantom phase in the present or future, which may give rise to a future singularity (see Ref.~\cite{Bamba:2011dk}). In the era of precise cosmology, a deeper study on the large number of dark energy models is an essential task in order to avoid a problem of degeneracy, since most of dark energy candidates reproduce the same cosmological evolution with no significant differences. In the case of modified gravity, some studies about the stability of the cosmological solutions \cite{delaCruzDombriz:2011wn}, or the exploration of cosmological perturbations \cite{Perturbations} and CMB anisotropies \cite{Bourhrous:2012kr}  provide some restrictions on the  gravitational action.  \\

In the present manuscript, the study of the so-called viable $f(R)$ gravity is explored by analyzing the corresponding cosmological evolution, and its scalar-tensor representation. It is well known that $f(R)$ gravity can be redefined as a kind of Brans-Dicke theory in terms of a scalar field with a null kinetic term (see Refs.~\cite{MyF(R)1,Kobayashi:2008wc} and references therein). Then, the study of the dynamics of the scalar-tensor equivalence is an important tool for a better understanding of $f(R)$ gravities, and the cosmological evolution under these theories. In this sense, an analysis of the scalar-tensor equations in vacuum for Friedmann-Lema\^itre-Robertson-Walker (FLRW) metrics is performed by studying the corresponding dynamical system. Then, the stability of the cosmological evolution is analyzed, as well as the occurrence of future singularities in viable $f(R)$ gravities. For these purposes, two characteristic models of the class of viable $f(R)$ theories are considered (see Refs.~\cite{f(R)viable1,f(R)viable2}), where the cross of the phantom barrier emerges as a common feature in both models. Then, the possibility of the occurrence of a future singularity is studied through the analysis of the phase space of the scalar-tensor counterpart. Moreover, the so-called {\it Little Rip},  a non singular super accelerated expansion phase,  is also explored. The {\it Little Rip} consists of a very strong accelerated expansion, whose strength may affect some bound systems, as planetary systems due to the Hubble parameter turns out a monotonically increasing function that diverges at infinity (see Ref.~\cite{LittleRip}). Previously, the {\it Little Rip} has been considered in several dark energy models \cite{Frampton:2011rh}, and also in the framework of modified gravities, \cite{Nojiri:2011kd}. In addition, even in the case that the Hubble parameter remains finite as time goes to infinity, the inertial force induced by the expansion may be strong enough to affect bound structures, producing a similar scenario called {\it Pseudo-Rip} \cite{Frampton:2011aa}. Here the occurrence of a {\it Little Rip} or a {\it Pseudo-Rip} is analyzed in the framework of viable $f(R)$ theories, in particular the effects on the Earth-Sun system are qualitatively studied.\\

The paper is organized as follows: section \ref{General} is devoted to the introduction of viable $f(R)$ theories and their main cosmological properties. In section \ref{Evolution}, the evolution of the effective EoS parameter of the universe is explored for two characteristic models. Section \ref{Scalar} deals with the analogous scalar-tensor theory for $f(R)$ gravity and the behavior of the phase space. Finally, in section \ref{Singularities},  the possibility of the presence of cosmological singularities is discussed as well as the occurrence of a Little Rip.

\section{Viable $f(R)$ theories}

\label{General}

Let's start by reviewing viable $f(R)$ theories in the cosmological context. The  corresponding gravitational action is given by,
\be
S=\int d^4x \sqrt{-g}\left[f(R)+2\kappa^2\mathcal{L}_m\right] \ .
\label{1.1}
\ee
Here the coupling constant is given by $\kappa^2=8\pi G$, whereas $\mathcal{L}_m$ stands for the Lagrangian corresponding to the matter content. Note that the  Hilbert-Einstein action is recovered for $f(R)=R$. 
Then, the field equations corresponding to the action (\ref{1.1}) are obtained by the variation of this action with respect to the metric tensor $g_{\mu\nu}$, 
\be
 R_{\mu\nu} f_R(R)- \frac{1}{2} g_{\mu\nu} f(R) + g_{\mu\nu}  \Box f_{R}(R) -  \nabla_{\mu} \nabla_{\nu}f_{R}(R)=\kappa^2T^{(m)}_{\mu\nu}\ .
\label{1.2}
\ee
Here the subscript $_{R}$ denotes derivatives with respect to $R$. We  assume a flat FLRW metric, $ds^2=-dt^2+a(t)^2\sum_{i=1}^3dx^{i2}$, so the modified FLRW equations are obtained through the  $00$ and $ij$ components of the field equations (\ref{1.2}),   
\bea
H^2=\frac{1}{3f_R}\left[\kappa^2 \rho_m +\frac{Rf_R-f}{2}-3H\dot{R}f_{RR}\right]\ , \nn
-3H^2-2\dot{H}=\frac{1}{f_R}\left[\kappa^2p_m+\dot{R}^2f_{RRR}+2H\dot{R}f_{RR}+\ddot{R}f_{RR}+\frac{1}{2}(f-Rf_R)\right]\ ,
\label{1.3}
\eea
where dots denote derivatives with respect to the cosmic time, the Hubble parameter is defined as usual $H(t)=\dot{a}/a$, and the Ricci scalar is $R=6\ (2H^2+\dot{H})$. Here, we are interested in studying a subclass of modified gravities, the  so-called viable $f(R)$ gravities, which are usually described by actions of the type, 
\be
f(R)=R+F(R)\ .
\label{1.4b}
\ee
This action basically represents the Hilbert-Einstein action plus an additional term that should have only effects at cosmological scales, while at local scales, GR should be recovered. In order to avoid serious problems with known physics, one has to choose the $F(R)$ function such that the theory contains flat solutions and passes also local gravity tests (see \cite{f(R)GRtests}). Viable $f(R)$ gravities are able to satisfy  these constraints, and avoid the anti-gravity regime by imposing positivity in the first derivative of the action, $1+F_R>0$, and by the appropriate form of the function $F(R)$ (see \cite{f(R)viable1}-\cite{f(R)viableOthers}). Note that both equations in (\ref{1.3}) are written in such a way that  $F(R)$-terms appear in the matter side (see Ref.~\cite{MyF(R)1}); thus we may define an energy density  for the extra $F(R)$-terms. Hence, the first FLRW equation in (\ref{1.3}) can be rewritten for the kind of actions (\ref{1.4b}) in terms of the cosmological parameters $\Omega_i$,
\[
1=\Omega_{m}+\Omega_{F(R)}\ , \quad where
\]
\be
\Omega_m=\frac{\rho_m}{\frac{3}{\kappa^2}H^2}\ , \quad \Omega_{F(R)}=\frac{1}{3H^2}\left(\frac{RF_R-F}{2}-3H\dot{R}F_{RR}-3H^2F_R\right)\ .
\label{1.4}
\ee
Note that the first Friedmann equation (\ref{1.3}) takes a simple form, with two fluids contributing to the scale factor dynamics. In addition, the continuity equation $\nabla_{\mu}T^{\mu\nu}=0$ for a perfect fluid with an EoS $p_m=w_m\rho_m$ is given by $\dot{\rho}_m+3H(1+w_m)\rho_m=0$. 
In absence of any kind of matter,  the dynamics of the universe are described solely by $F(R)$-component, which may be chosen so that  reproduces (or at least contributes) to the early inflationary phase and  the late-time accelerating epoch.  To reproduce the whole history of the universe, the following conditions on the $F(R)$ function have been proposed (see \cite{f(R)viable2}):
\begin{enumerate}
\renewcommand{\labelenumi}{\roman{enumi}}
\item) Accelerating expansion during inflation may occur  under one of the following conditions:
\be
\lim_{R\rightarrow\infty} F(R)\rightarrow-\Lambda_i\ \quad or \quad \lim_{R\rightarrow\infty} F(R)\rightarrow  \alpha R^n\ .
\label{1.7}
\ee
In the first situation in (\ref{1.7}),  $F(R)$ function behaves as an effective cosmological constant at early times, while the second condition yields also an accelerating expansion where the scale factor behaves as $a(t)\sim t^{m}$, with $m=\frac{2 n^2+1-3 n}{2-n}$ (see Ref.~\cite{Goheer:2009ss} and references therein).
\item) In order to reproduce late-time acceleration, we can impose a similar condition on  $F(R)$. In this case, Ricci scalar has a finite value, which is assumed to be the current one, so that the condition is expressed as
\be
F(R_0)=-2R_0 \quad F'(R_0)\sim 0\ .
\label{1.8}
\ee
This basically means that  extra terms in the action behave as an effective cosmological constant at the present time, $p\sim-\rho$. However, as the effective fluid, coming from extra terms in the action, is not exactly a cosmological constant, the evolution would be quite different, since the universe may enter in a phantom phase \cite{Bamba:2011dk}, or may produce oscillations along the cosmological evolution \cite{Bamba:2010iy,Elizalde:2011ds,Appleby:2009uf}. We will explore the evolution, and the crossing of the phantom barrier in viable $f(R)$ gravity in the next sections.
\end{enumerate}
Along this manuscript, we will focus on the study of two characteristic models of this kind of viable modified gravity, proposed by Hu and Sawicki in Ref.~\cite{f(R)viable1}, and by Nojiri and Odintsov in Ref.~\cite{f(R)viable2},  whose actions are given by,
\be
F_{HS}(R)=-R_{HS}\frac{c_1(R/R_{HS})^n}{c_2(R/R_{HS})^n+1}\ , \quad F_{NO}(R)=\frac{R^n(a R^n-b)}{1+c R^n}\ .
\label{1.8a}
\ee
Here, $\{c_1, c_2,n\}$ are free parameters and $R_{HS}=\kappa^2\rho^0_m$ according to Ref.~\cite{f(R)viable1}, while $\{a,b,c,n\}$ are free parameters for the second model in (\ref{1.8a}). The first model in (\ref{1.8a}) has been widely studied in Refs.~\cite{f(R)deSitter,Bamba:2010iy,Appleby:2009uf}, where was proven that the universe evolution can go through several de Sitter points, being some of them stable and  other unstable, which may be identified  to the current accelerated era and to the inflationary epoch. Similarly, the second model in(\ref{1.8a}) was studied in Ref.~\cite{Saez-Gomez1}, also with the presence of an extra field, and was shown that can well reproduce both accelerated epochs, mimicking a cosmological constant at the present. In the next sections, we will study the evolution of both models, focusing on the possibility of the crossing of the phantom barrier, and its consequences. 

\section{Cosmological evolution in viable $f(R)$ gravity}
\label{Evolution}

In this section, we explore the cosmological evolution for the models considered above. For that reason, it is  convenient to express the equations (\ref{1.3}) in terms of the redshift $1+z=\frac{a_0}{a(t)}$ instead of the cosmic time $t$, where $a_0$ is the value of the scale factor at the present time $t_0$, such that the current epoch corresponds to $z=0$ Then, the time derivative is transformed as $\frac{d}{dt}=-(1+z)H\frac{d}{dz}$, and  the first FLRW equation in (\ref{1.3}) and the continuity equation yield,
\be
H^2(z)=\frac{1}{3f_R}\left[\kappa^2\rho_m(z)+\frac{R(z)f_R-f}{2}+3(1+z)H^2f_{RR}R'(z)\right]\ ,
\label{2.2}
\ee
\be
(1+z)\rho_m'(z)-3(1+w_m)\rho_m(z)=0\ ,
\label{2.3}
\ee
where now primes denote derivatives with respect to the redshift. Then, the Ricci scalar can be rewritten as $R=6\left[2H^2(z)-(1+z)H(z)H'(z)\right]$, while the equation (\ref{2.3}) is easily solved for a constant EoS parameter $w_m$,
\be
\rho(z)=\rho_0(1+z)^{3(1+w_m)}\ .
\label{2.4}
\ee
Here $\rho_0$ is the value of the matter energy density at the present epoch $z=0$, which can be rewritten as $\rho_0=\frac{3}{\kappa^2}H_0^2\Omega_m^0$. Then, we can fit the current values of the cosmological parameters using the observational data \cite{WmapData}, where $H_0=100\, h \ km\ s^{-1}\ Mpc^{-1}$ with $h=0.71\pm 0.03$ and the matter density $\Omega_m^0=0.27 \pm 0.04$, whereas EoS  is given by $w_m=0$ (cold dark matter and baryons). The equation (\ref{2.2}) is a second order differential equation on $H(z)$, so by fixing the initial conditions,  the corresponding cosmological evolution can be obtained through the Hubble parameter in terms of the redshift  and the future evolution (with $-1<z<0$) might be explored. In order to illustrate the cosmological evolution, let us consider the effective equation of state parameter $w_{eff}$, which is defined as
\be
w_{eff}=\frac{p_{F(R)}+p_m}{\rho_{F(R)}+\rho_m}=-1-\frac{2\dot{H}(t)}{3H^2(t)}=-1+\frac{2(1+z)H'(z)}{3H(z)}\ .
\label{2.5}
\ee
For simplicity we redefine the Hubble parameter as $H(z)=H_0\ h(z)$, such that $h(0)=1$, whereas the initial condition on the first derivative $h'(0)$ can be fixed by extrapolating a particular model at $z=0$. In this sense, by  assuming  $\Lambda$CDM model as a good approximation at $z=0$,  FLRW equations are given by, 
\be
H^2=\frac{\kappa^2}{3}\rho_m+\frac{\Lambda}{3}\ , \quad \dot{H}=-\frac{\kappa^2}{2}\rho_m\ .
\label{2.6}
\ee
By the change of variables $t\rightarrow z$, and $H(z)=H_0\ h(z)$, the second equation in (\ref{2.6}) at $z=0$ yields,  
\be
h'(z)=\frac{\kappa^2}{2}\frac{\rho_m}{H^2_0\ (1+z)h(z)}\ , \quad \rightarrow \quad h'(0)=\frac{\kappa^2}{2H_0^2}\rho_m^0=\frac{3}{2}\Omega_m^0\ .
\label{2.7}
\ee
Hence, the cosmological evolution of the $f(R)$ models (\ref{1.8a}) is explored by assuming that the universe mimics $\Lambda$CDM model (\ref{2.6}) at present time ($z=0$), i.e. by imposing the initial conditions (\ref{2.7}). In addition,  the assumption that  the universe has already crossed (or will do very soon) the phantom barrier at $z=0$ is also explored and compared with the other conditions. For this last case, the first derivative of the Hubble parameter with respect to the redshift should be approximately zero or negative, $h'(0)\lesssim 0$. \\

\begin{figure*}[h!]
	\centering
\subfloat[]{\includegraphics[width=0.40\textwidth]{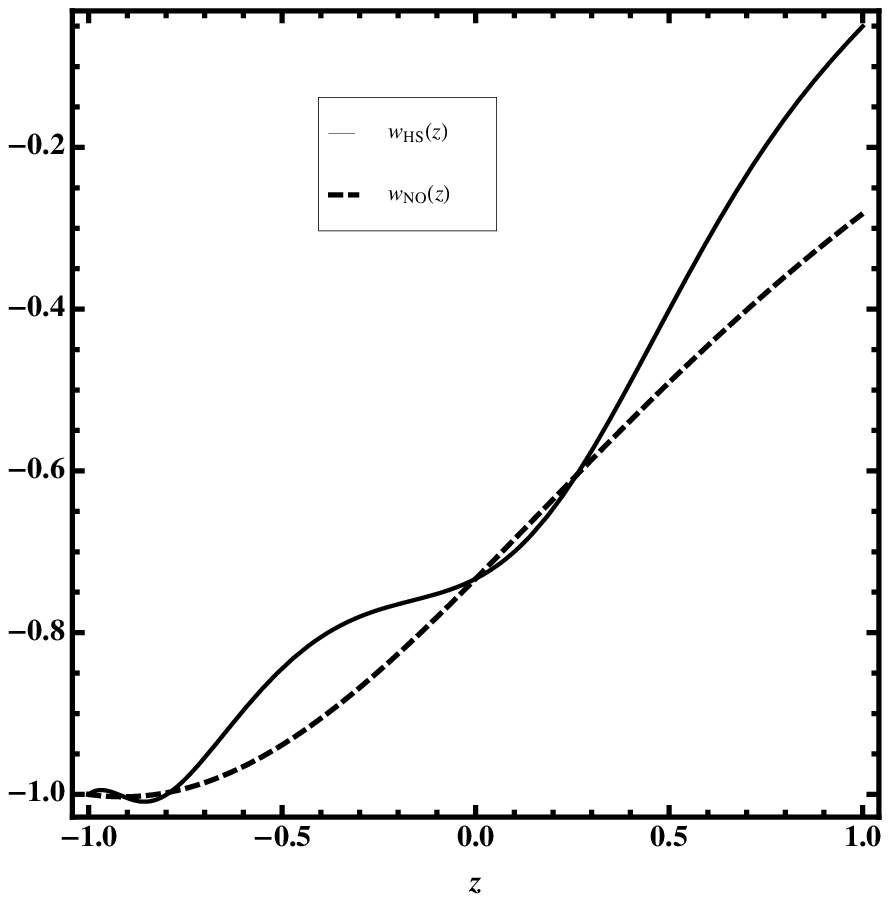}\label{fig1a}}\, \, \, \, \, \,
\subfloat[]{\includegraphics[width=0.40\textwidth]{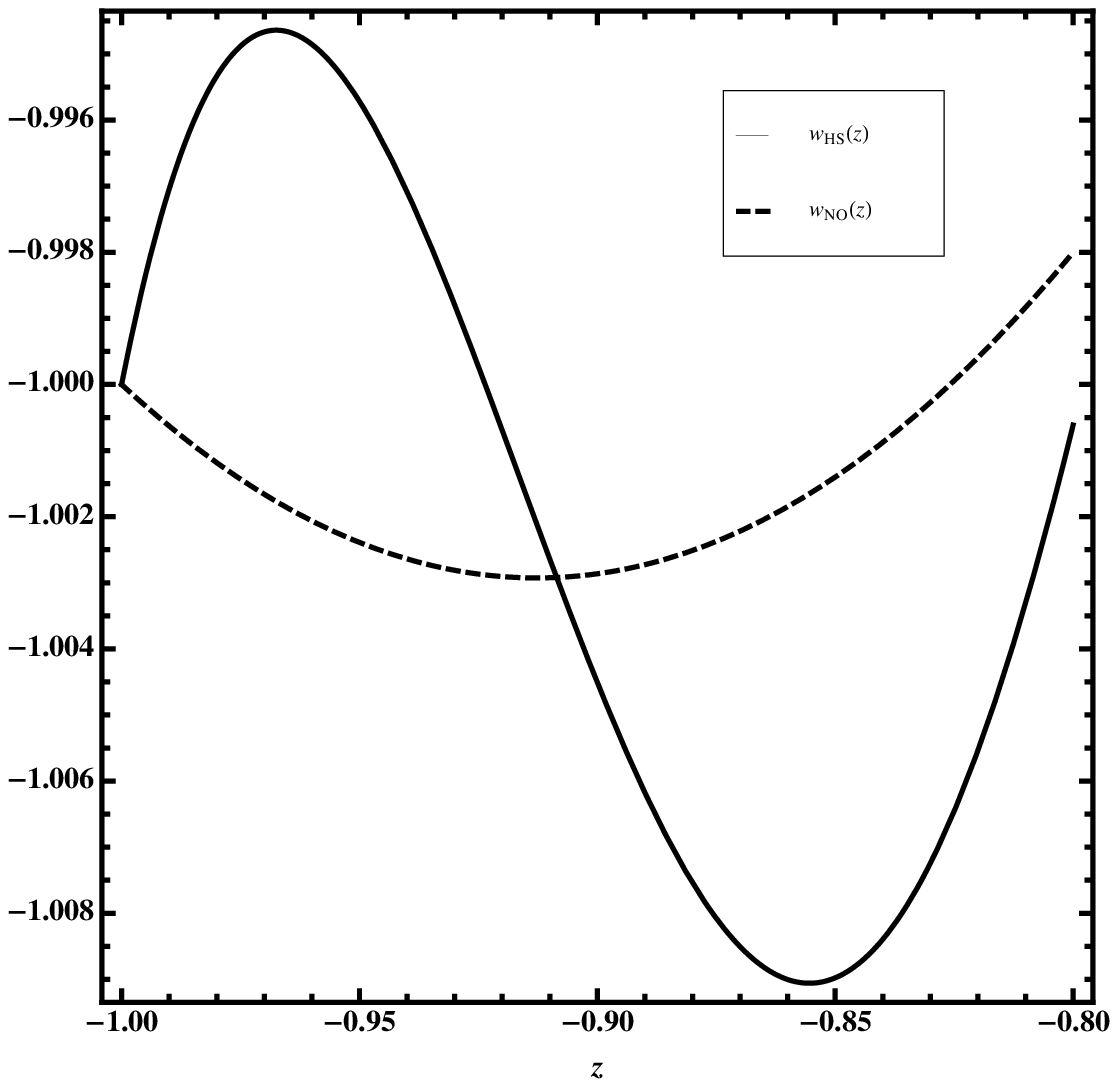}\label{fig1b}}
 \caption{\footnotesize{Evolution of the EoS parameter (\ref{2.5}) as a function of the redshift $z$ for the models $F_{HS}(R)$ and $F_{NO}(R)$. Here we have assumed the initial conditions $h(0)=1$ and $h'(0)=0.4$, according to (\ref{2.7}). The panel \ref{fig1a} shows the evolution from the past to the future, where the phantom barrier is crossed for negative redshifts in both models. The panel~\ref{fig1b} shows in more detail the range $-1<z<-0.8$.}}
\end{figure*}

Let's now analyze the behavior of the $f(R)$ models (\ref{1.8a}). For illustrative proposes we assume here $c_1=2$ and $c_2=1$, which are dimensionless parameters, and a power of $n=1$ for the Hu-Sawicki action, whereas  $n=1$, $a=0.1/H_0^2$ , $b=1$, $c=0.05/H_0^2$ are set for the Nojiri-Odintsov model. In Fig.~1, the evolution of the EoS parameter (\ref{2.5}) is depicted with respect to the redshift by assuming the initial conditions (\ref{2.7}) at $z=0$. In spite of the evolution of the EoS parameter is quite similar to the $\Lambda$CDM model for positive redshifts, the universe enters in a phantom phase along the future (negative redshifts), where the EoS parameter turns out less than $-1$ in both models, and oscillates close to $z=-1$ for the Hu-Sawicki model as shown in Fig.~\ref{fig1b} in more detail. In Fig.~2, the EoS parameter (\ref{2.5}) is also evaluated, but  assuming a phantom transition at $z=0$. The evolution in this second case starts to oscillate before in both models, and with larger amplitude than in Fig.~1. Furthermore, the HS model presents oscillations in both cases, with a larger amplitude in Fig.~2, while the oscillations in the NO model are only identified in Fig.~2. This is even more clear in Fig.~3, where the deceleration parameter $q=-a\ddot{a}/\dot{a}^2$ is depicted. Fig.~\ref{figQa} corresponds to $\Lambda$CDM initial conditions, where the deceleration parameter presents a similar behavior to $\Lambda$CDM model, whereas Fig.~\ref{figQb}  shows a clear deviation with strong oscillations in comparison with the smooth curve of  $\Lambda$CDM model at small redshifts. In any case, the value $w=-1$ seems to be an asymptotic stable point in both models. Note that this fact is very common  in $f(R)$ gravity, since the study of the dynamical system of the set of equations (\ref{1.3}) reveals that $\dot{H}=0$ is a critical point, which may lead to an asymptotic stable EoS $w_{eff}=-1$ (see Ref.~\cite{f(R)deSitter}). This issue will be discussed in more detail below by analyzing the scalar-tensor counterpart, where is found that dS solution is an asymptotic stable point in both models for the conditions studied here.
\begin{figure*}[h!]
	\centering
\subfloat[]{\includegraphics[width=0.40\textwidth]{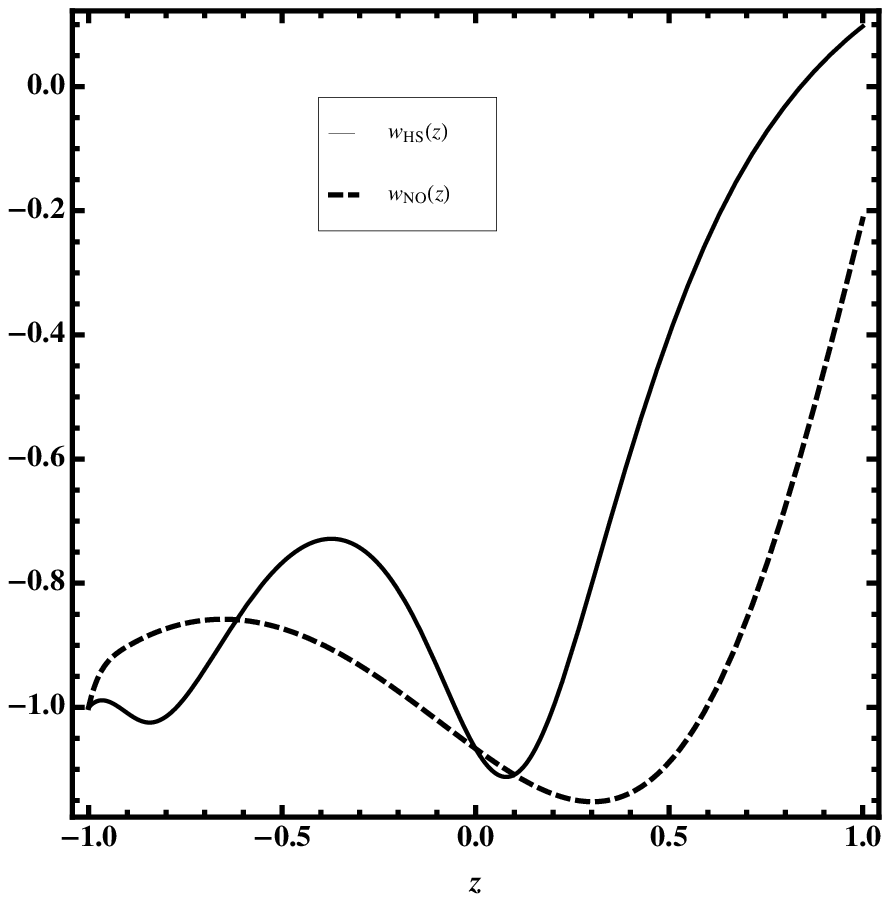}\label{fig2a}}\, \, \, \, \, \,
\subfloat[]{\includegraphics[width=0.40\textwidth]{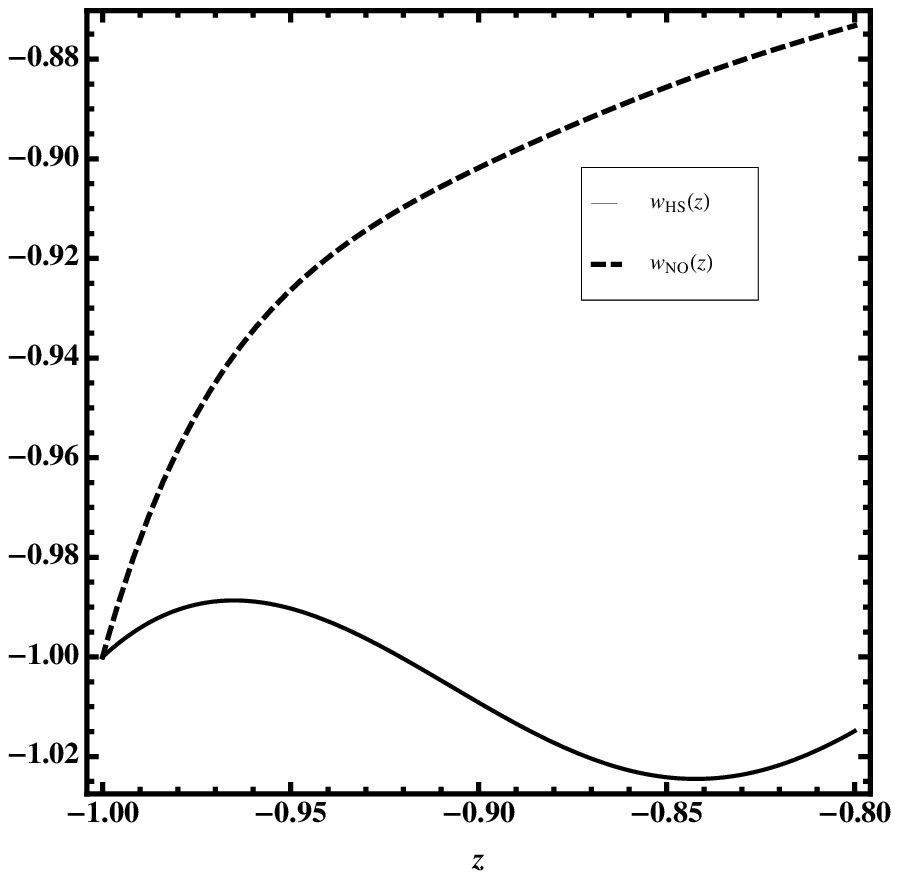}\label{fig2b}}
 \caption{\footnotesize{Evolution of the EoS parameter (\ref{2.5}) as a function of the redshift $z$ for the models $F_{HS}(R)$ and $F_{NO}(R)$ and the initial conditions $h(0)=1$ and $h'(0)=-0.1$. As shown in  panel \ref{fig2a}, the EoS crosses the phantom barrier  before $z=0$ in both models (as natural due according to the initial conditions), while it oscillates along the barrier in the future. In more detail,  the future evolution is shown in the Fig.~\ref{fig2b}. Both models tend to $w=-1$ when approaching $z=-1$.}}
\end{figure*}
\begin{figure*}[h!]
	\centering
\subfloat[]{\includegraphics[width=0.40\textwidth]{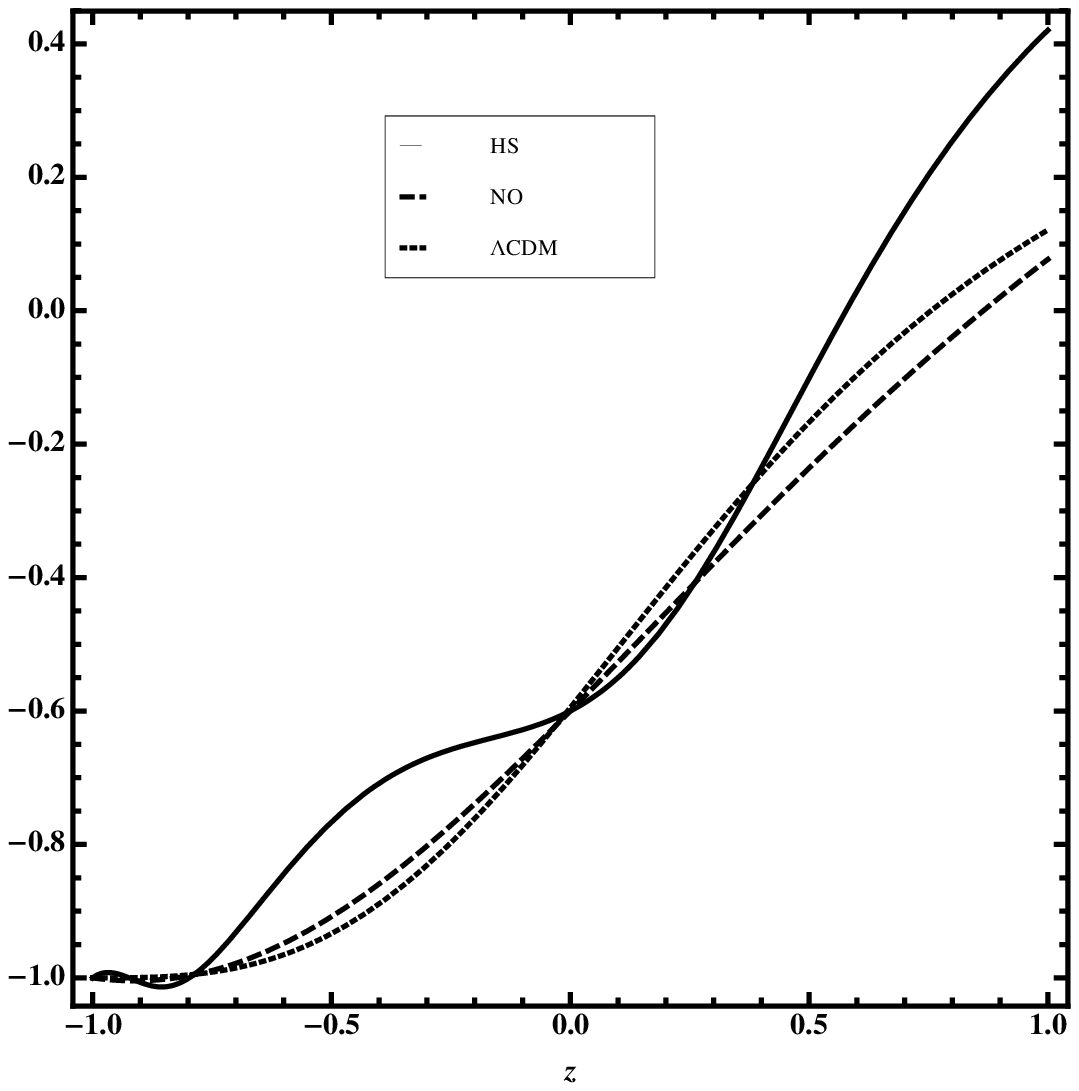}\label{figQa}}\, \, \, \, \, \,
\subfloat[]{\includegraphics[width=0.40\textwidth]{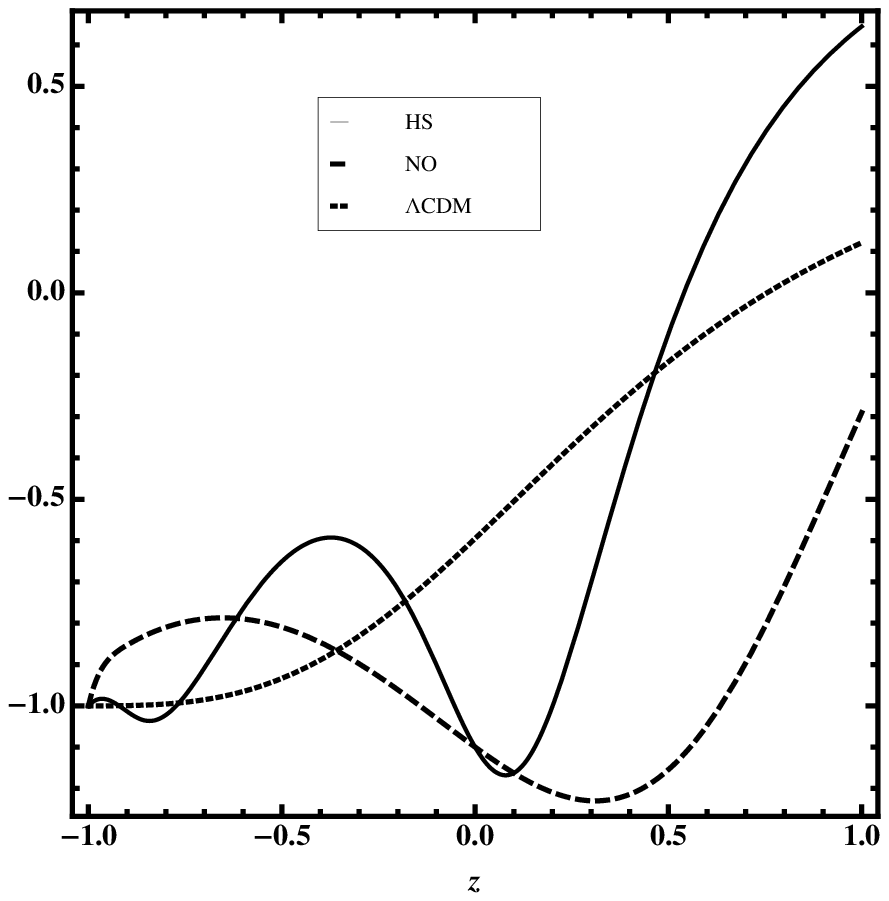}\label{figQb}}
 \caption{\footnotesize{Evolution of the deceleration parameter $q=-a\ddot{a}/\dot{a}^2$ for the models $F_{HS}(R)$ and $F_{NO}(R)$ in comparison with the $\Lambda$CDM model. The panel \ref{figQa} corresponds to the  initial conditions $h(0)=1$ and $h'(0)=0.4$, while the panel \ref{figQb} corresponds to $h(0)=1$ and $h'(0)=-0.1$.}}
\end{figure*}

\begin{figure*}[h!]
	\centering
\subfloat[]{\includegraphics[width=0.40\textwidth]{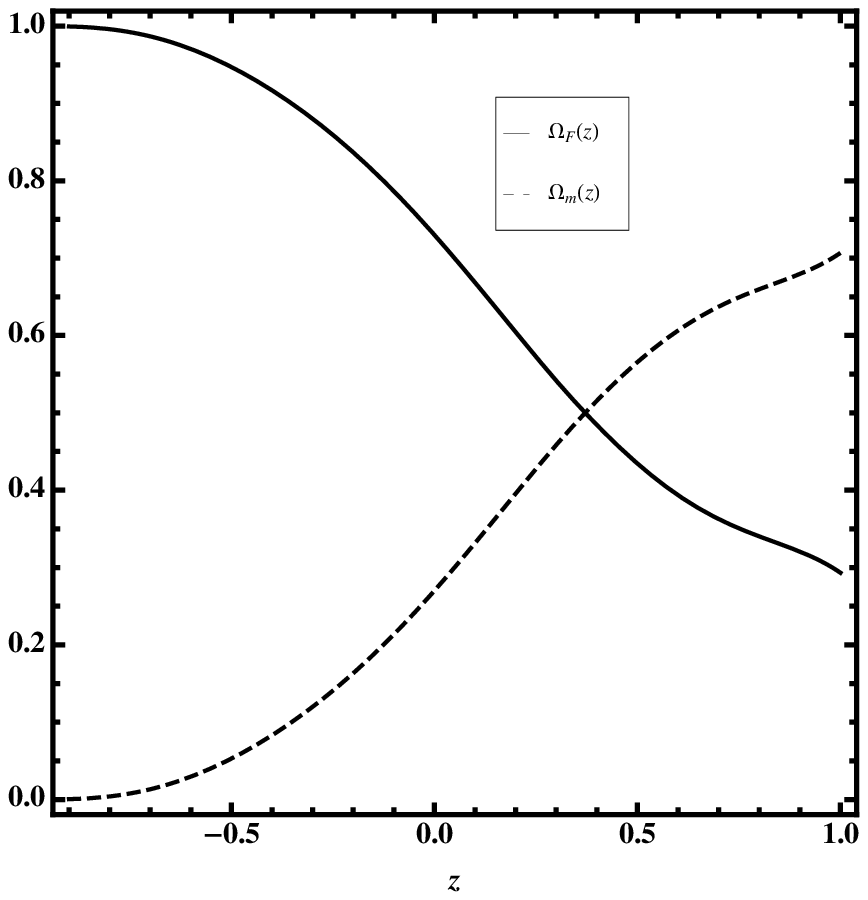}\label{fig5a}}\, \, \, \, \, \,
\subfloat[]{\includegraphics[width=0.40\textwidth]{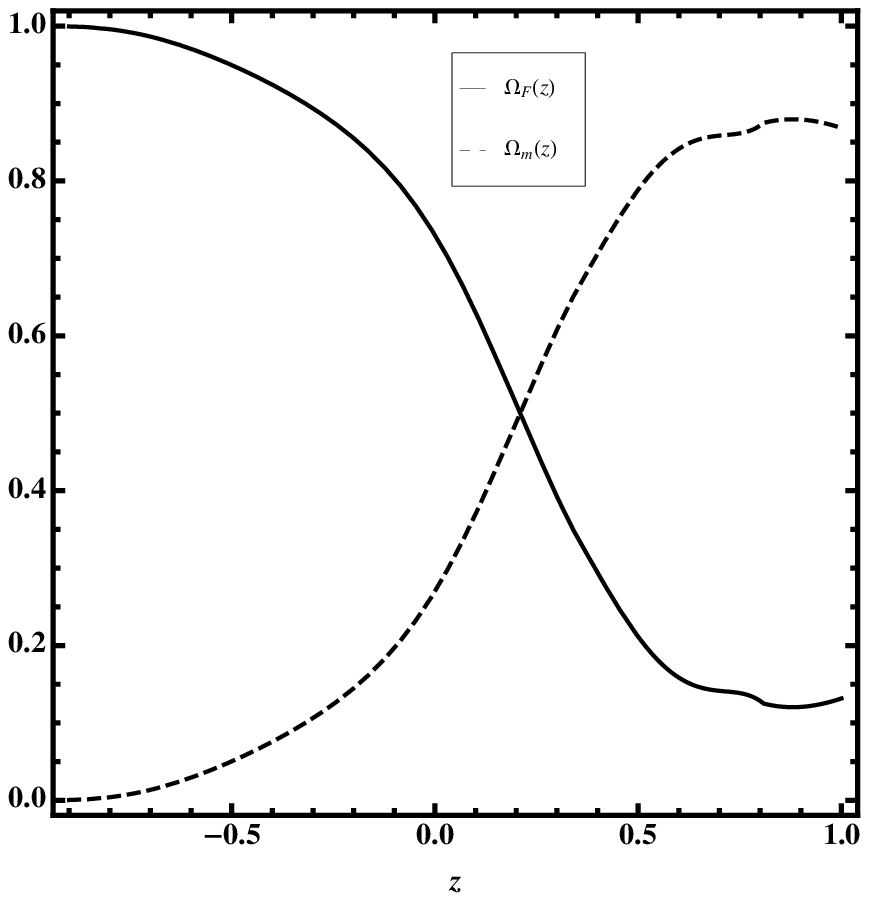}\label{fig5b}}
 \caption{\footnotesize{Evolution of the cosmological parameters $\{\Omega_m,\Omega_{F}\}$ defined in (\ref{1.4}) in the framework of the HS model, and assuming the initial conditions, $h(0)=1$ and $h'(0)=0.4$ (Fig.~\ref{fig5a}), and $h(0)=1$ and $h'(0)=-0.1$ (\ref{fig5b}). Both plots are very similar, but the panel \ref{fig5b} shows that the cosmological parameter $\Omega_F$ decays for $0<z<1$, and starts to increase again at $z\sim1$, an anomalous behavior.}}
\end{figure*}
\begin{figure*}[h!]
	\centering
		\subfloat[]{\includegraphics[width=0.40\textwidth]{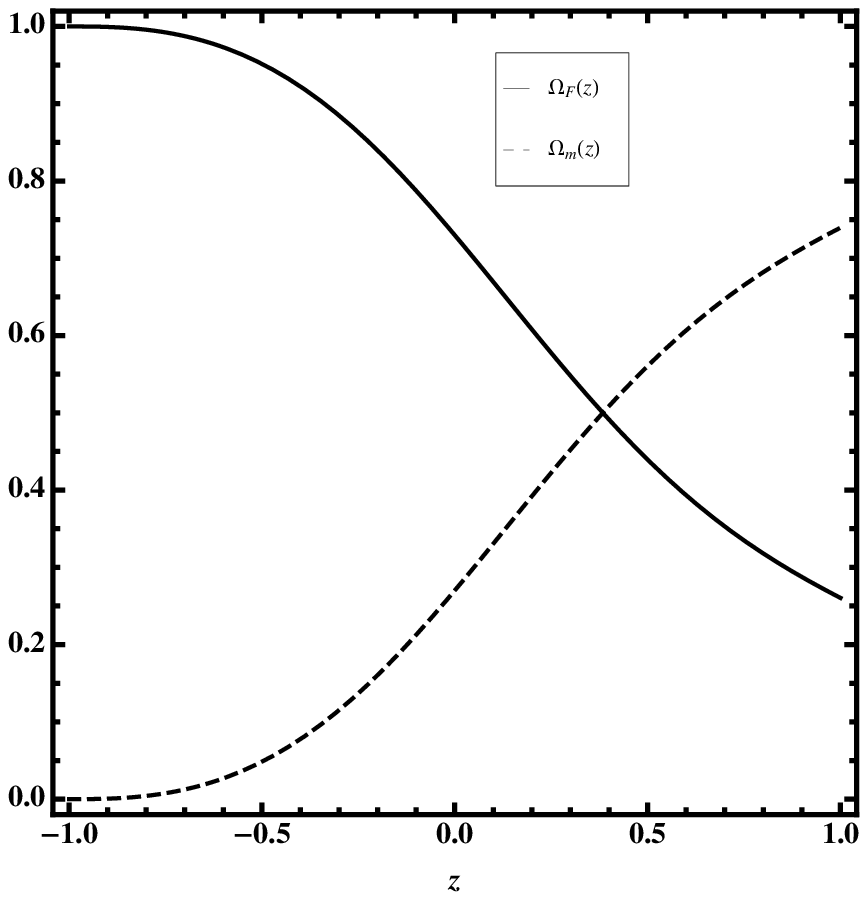}\label{fig6a}}\, \, \, \, \, \,
\subfloat[]{\includegraphics[width=0.40\textwidth]{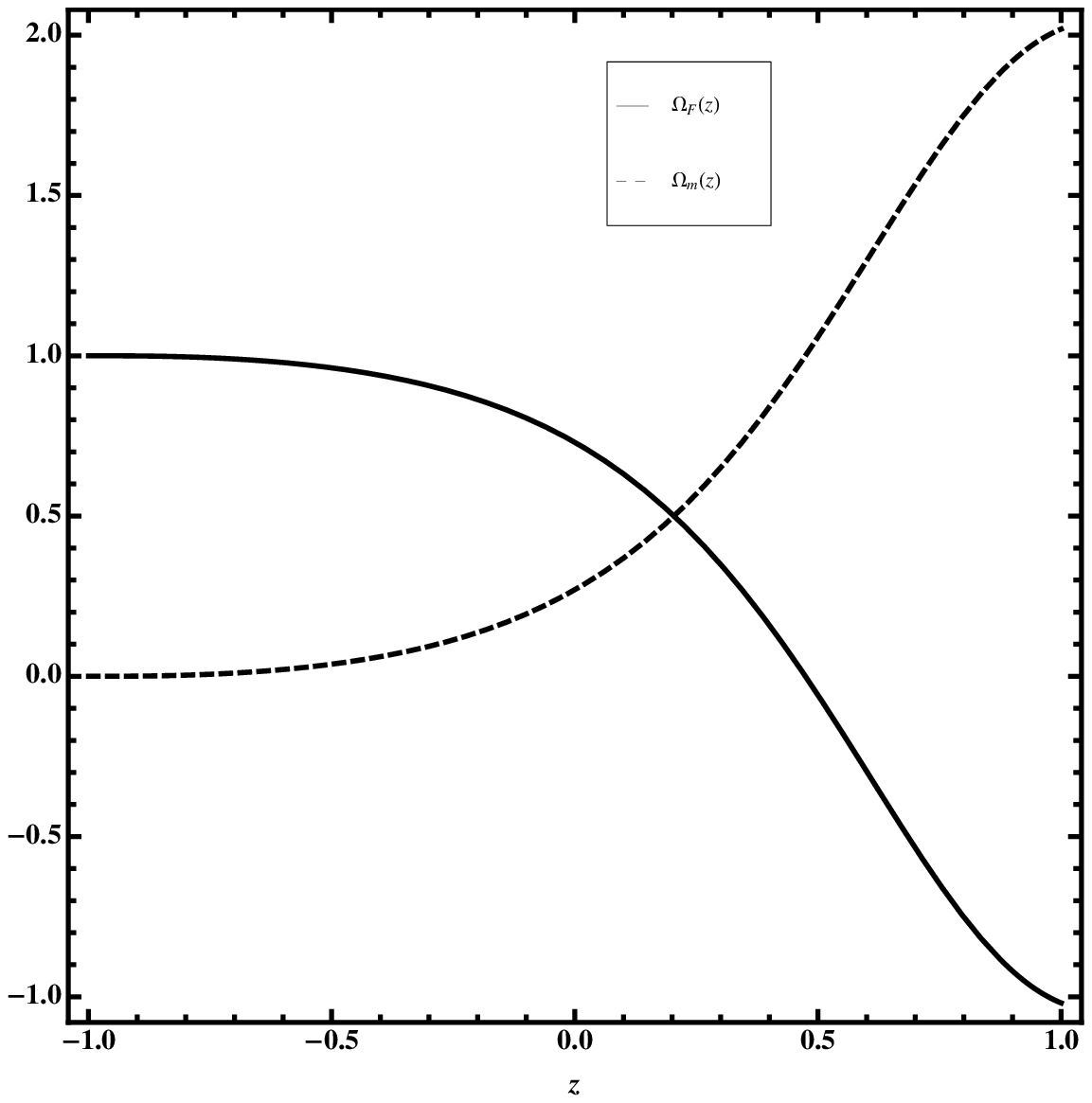}\label{fig6b}}
 \caption{\footnotesize{Evolution of the cosmological parameters $\{\Omega_m(z),\Omega_{F}(z)\}$  in the $F_{NO}(R)$: panel \ref{fig6a} corresponds to $h(0)=1$ and $h'(0)=0.4$, while panel \ref{fig6b} represents the evolution for the conditions $h(0)=1$ and $h'(0)=-0.1$. In Fig.~\ref{fig6b}, the growth of matter density $\Omega_m$ experiences a strong increase for positive redshifts.}}
\end{figure*}


In addition, figures 4 and 5 show the evolution of the cosmological parameters $\{\Omega_m,\Omega_F\}$ with respect to the redshift for the HS and NO models respectively. Note that the choice of the initial conditions affect the cosmological evolution in both models. In the HS model, both panels in Fig.~4 present the same evolution for negative redshifts, whereas the evolution deviates along positive redshifts, as shown in Fig.~\ref{fig5b}. In the NO model, when phantom initial conditions are assumed, $\Omega_F$ becomes negative at positive redshifts, fig.~\ref{fig6b}.  \\

Note that the viable modified models considered here (\ref{1.8a}) produce some oscillations along the cosmological evolution, a fact pointed out before for this kind of models in Ref.~\cite{Bamba:2010iy,Elizalde:2011ds}. Furthermore, the transition to the phantom epoch occurs in general for models (\ref{1.8a}). Moreover, the election of phantom conditions at $z=0$ gives an anomalous behavior of the cosmological parameters for both models when $z>0$. 


\section{Scalar-tensor representation of $f(R)$ gravity}
\label{Scalar}

At this stage, let us consider the equivalent scalar-tensor theory of $f(R)$ gravity (see for example, \cite{MyF(R)1,Kobayashi:2008wc} and references therein),
 \be
S=\int d^4x \sqrt{-g}\left[\phi\ R-V(\phi)+2\kappa^2\mathcal{L}_m\right] \ .
\label{3.1}
\ee
By varying the action (\ref{3.1}) with respect to the scalar field $\phi$, the relation between both theories is obtained, and the $f(R)$ action (\ref{1.1}) is recovered,
\be
R=V'(\phi) \quad \rightarrow \quad \phi=\phi(R)\ , \\
\Rightarrow \quad f(R)=\phi(R)R-V(\phi(R))\ .
\label{3.2}
\ee
While the scalar field and the potential are related to a particular $f(R)$ by, 
\be
\phi=f_R(R)\ , \quad V(\phi(R))=f_R(R)\ R-f(R)\ .
\label{3.3}
\ee
Hence, it is straightforward to reconstruct the equivalent scalar-tensor theory (\ref{3.1}) for a particular $f(R)$ action. Moreover, by varying the action (\ref{3.1}) with respect to the metric $g_{\mu\nu}$, the field equations are obtained,
\[
R_{\mu\nu}-\frac{1}{2}g_{\mu\nu}R=\frac{\kappa^2}{\phi}T_{\mu\nu}^{(m)}+\frac{1}{\phi}(\nabla_{\mu}\nabla_{\nu}\phi-g_{\mu\nu}\Box\phi)-\frac{1}{2}g_{\mu\nu}V(\phi)\ ,
\]
\be
3\Box\phi=\kappa^2T^{(m)}+\phi\frac{dV(\phi)}{d\phi}-2V(\phi)\ ,
\label{3.4}
\ee
where the second equation is the trace of the field equations (\ref{3.4}). Hence, for a flat FLRW metric, the Friedmann equations yield,
\[
3H^2=\frac{1}{\phi}\left(\kappa^2\rho_m-3H\dot{\phi}+\frac{1}{2}V(\phi)\right)\ , \] 
\be
-3H^2-2\dot{H}=\frac{1}{\phi}\left(\kappa^2 p_m+\ddot{\phi}+2H\dot{\phi}-\frac{1}{2}V(\phi)\right)\ .
\label{3.5}
\ee
while the trace equation (\ref{3.4}) is given by,
\be
3\ddot{\phi}=\kappa^2(\rho_m-p_m)-\phi V'+2V-9H\dot{\phi}\ .
\label{3.6}
\ee
Let's reconstruct the corresponding scalar-tensor theory for the particular models considered in the previous sections. By the expressions (\ref{3.3}), the corresponding relation $\phi(R)$ for the models (\ref{1.8a}) is obtained,
\be
\phi_{HS}=1+\frac{c_1 c_2 R}{R_{HS}\left(1+\frac{c_2 R}{R_{HS}}\right)^2}+\frac{c_1}{1+\frac{c_2 R}{R_{HS}}}\ , \quad \phi_{NO}=1-\frac{c R(aR-b)}{(1+c R)^2}+\frac{a R}{1+c R}+\frac{aR-b}{1+cR}\ ,
\label{3.7}
\ee
where we have assumed a power of $n=1$ in both models (\ref{1.8a}). Then,  the scalar potential (\ref{3.3}) yields,
\[
V_{HS}(\phi)=\frac{1+c_1-\phi\pm2\ c_2\sqrt{c_1\ (1-\phi)}}{c_2}R_{HS}\ , \] 
\be
V_{NO}(\phi)=\frac{2\ a+c\ (1+b-\phi)\pm2\ \sqrt{(a+b\ c)\ (a+c-\phi)}}{c^2}\ .
\label{3.8}
\ee
Hence, the scalar-tensor representation for both models is not uniquely defined, but the potentials exhibit two branches due to the particular expressions of the $f(R)$ action considered here. In addition, both models introduce a boundary condition on the value of the scalar field, being $\phi<1$ in  the HS model, and $\phi<a+c/c$ in the NO model, a limit where both branches of the scalar potentials converge, as shown in Fig.~6. \\
\begin{figure*}[h!]
	\centering
\subfloat[]{\includegraphics[width=0.40\textwidth]{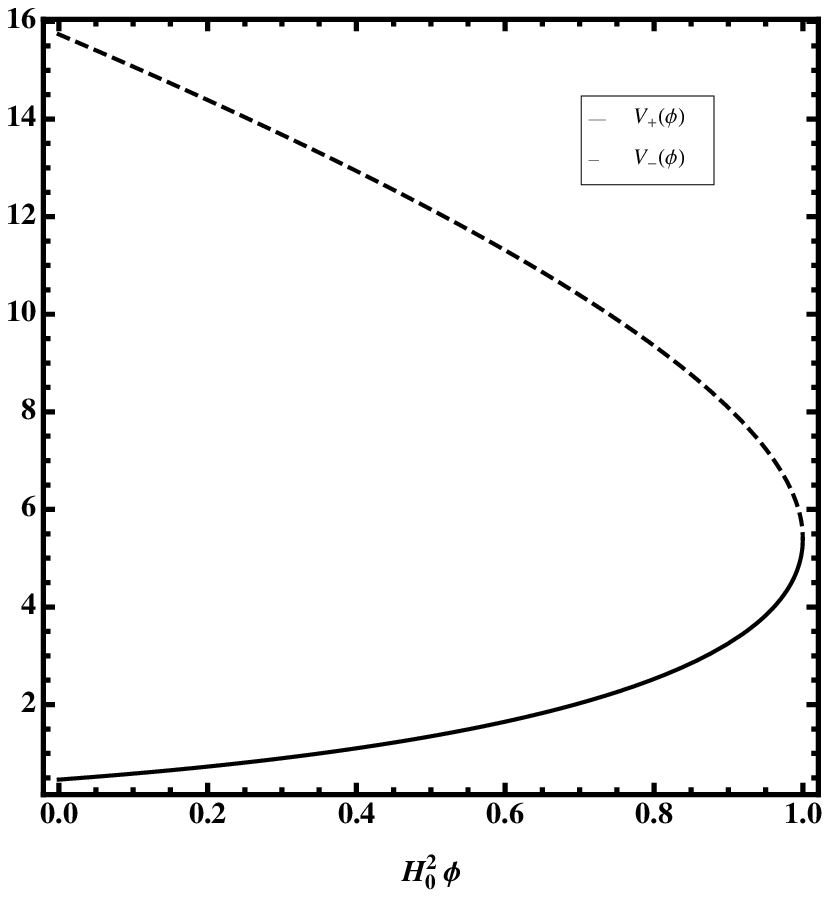}\label{fig7a}}\, \, \, \, \, \,
\subfloat[]{\includegraphics[width=0.40\textwidth]{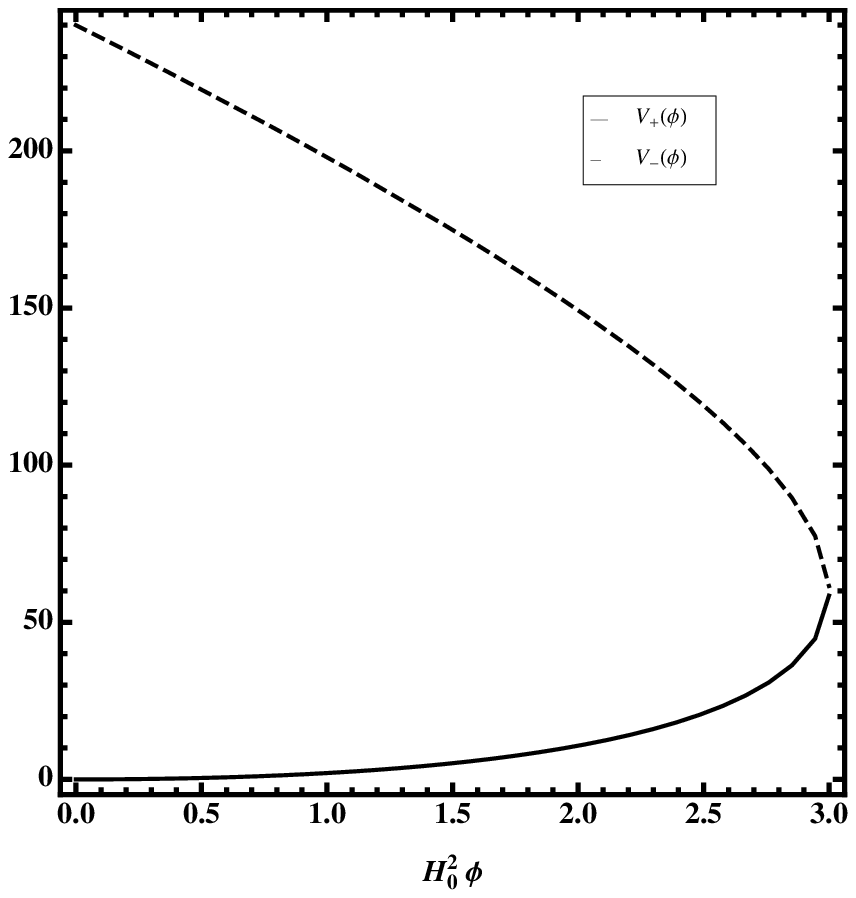}\label{fig7b}}
 \caption{\footnotesize{Scalar potentials $V(\phi)$ in terms of $H_0^2$,  for the HS model with $n=1$, $c_1=2$ , $c_2=1$ (panel \ref{fig7a}), and the NO model where $n=1$, $a=0.1/H_0^2$ , $b=1$, $c=0.05/H_0^2$, Fig.~\ref{fig7b}. Both potentials are not uniquely defined, but both branches converge to the same boundary of the scalar field. }}
\end{figure*}
Moreover, the evolution of the scalar field, $\phi(z)=f_R(R(z))$, shown in Fig.~7 is directly correlated with the behavior of the EoS parameter for negative redshifts, since the effective EoS parameter can be expressed now in terms of the scalar field as follows,
\be
w_{eff}=-1-\frac{\ddot{\phi}-H\dot{\phi}+\kappa^2\rho_m(1+w_m)}{\frac{1}{2}V(\phi)-3H\dot{\phi}+\kappa^2\rho_m}\ .
\label{3.8a}
\ee
At negative redshifts, the dust matter density is negligible, $\rho_m\propto (1+z)^3\sim0$, and the scalar field dominates, whereas the EoS yields,
\be
w_{eff}\sim-1-\frac{\ddot{\phi}-H\dot{\phi}}{\frac{1}{2}V(\phi)-3H\dot{\phi}}\ .
\label{3.8a}
\ee
Then, the behavior of the EoS parameter for small and negative redshifts is a direct consequence of the scalar field evolution, as observed by comparing Figs.~1 and 7. In the HS model, the oscillations of the scalar field, shown in Fig.~\ref{fig8a} are also observed in the evolution of the EoS parameter in Fig.~1, whereas the EoS parameter in the NO model describes a smooth curve in Fig.~1 that corresponds with the one of the scalar field in Fig.~\ref{fig8b}.\\
\begin{figure*}[h!]
	\centering
\subfloat[]{\includegraphics[width=0.40\textwidth]{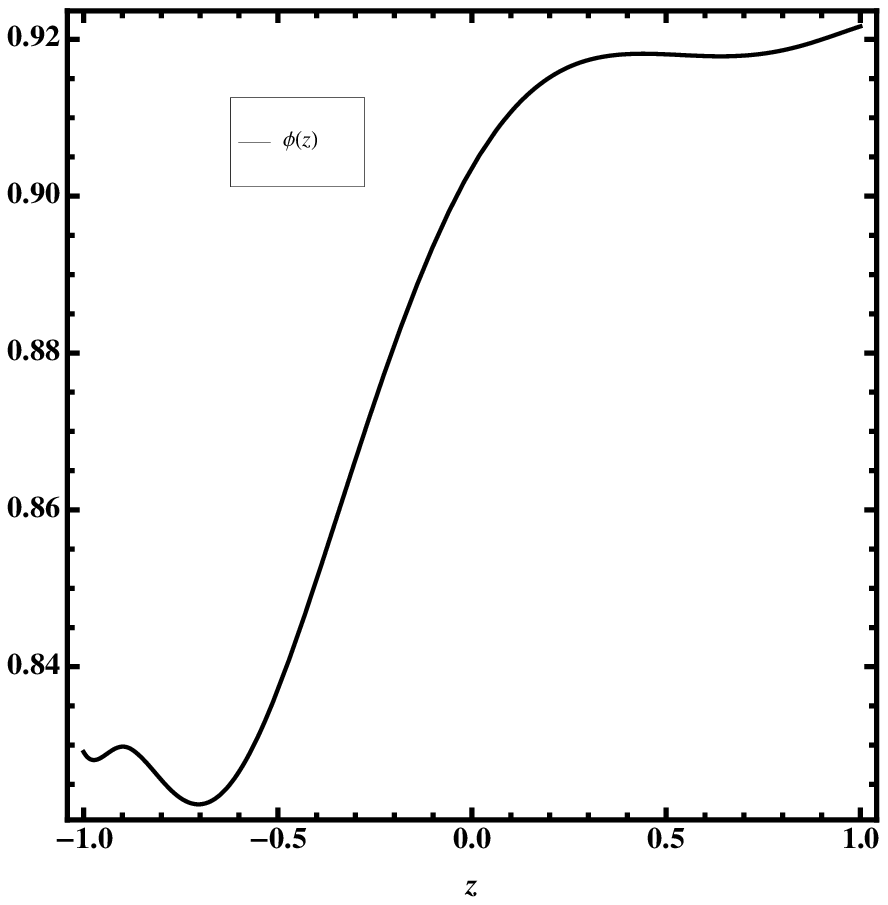}\label{fig8a}}\, \, \, \, \, \,
\subfloat[]{\includegraphics[width=0.40\textwidth]{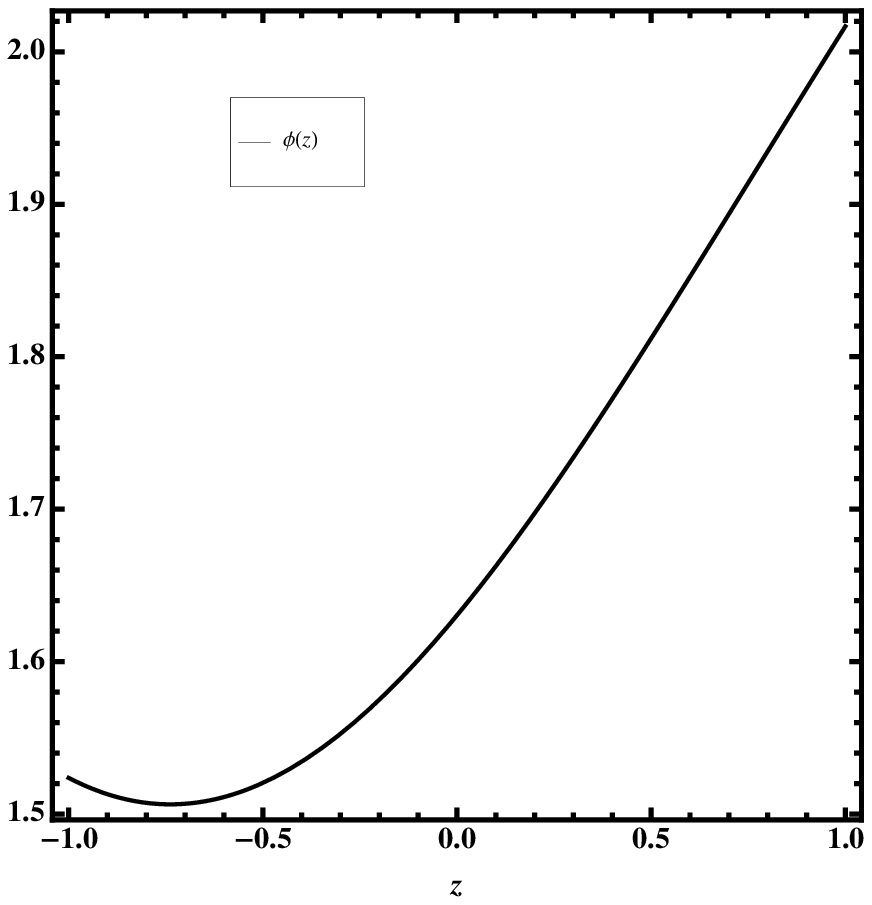}\label{fig8b}}
 \caption{\footnotesize{Evolution of the scalar field $\phi(z)$ for the HS model (\ref{fig8a}) and  the NO model (\ref{fig8b}). In both cases, the initial conditions are: $h(0)=1$, and $h'(0)=3\Omega_m^0/2$.}}
\end{figure*}
Let us now analyze the phase space $\{H,\phi\}$ in this kind of viable $f(R)$ gravity. For simplicity, we assume the FLRW equations (\ref{3.5}) in vacuum, so by combining the equations (\ref{3.5})-(\ref{3.6}), the following system results,
\be
\dot{H}=-2H^2+\frac{1}{6}V'(\phi)\ , \quad \dot{\phi}=\frac{1}{3H}\left[-3H^2\phi+\frac{1}{2}V(\phi)\right]\ .
\label{3.9}
\ee
Note that this is a dynamical system whose critical points are given by $\dot{H}=\dot{\phi}=0$, which corresponds to de Sitter solutions, as mentioned in the previous sections as a feature of $f(R)$ gravity. Then, the critical points are the solutions of  the algebraic equation,
\be
\phi_cV'(\phi_c)-2V(\phi_c)=0\ , \quad H_c=\sqrt{\frac{V'(\phi_c)}{12}}\ .
\label{3.9a}
\ee
In addition, by combining both equations in (\ref{3.9}), the equation of the phase space is obtained,
\be
\frac{dH}{d\phi}=\frac{\frac{V'(\phi)}{6}-2H^2}{\frac{V(\phi)}{6H}-H\phi}\ .
\label{3.10}
\ee
\begin{figure*}[h!]
	\centering
		\subfloat[]{\includegraphics[width=0.40\textwidth]{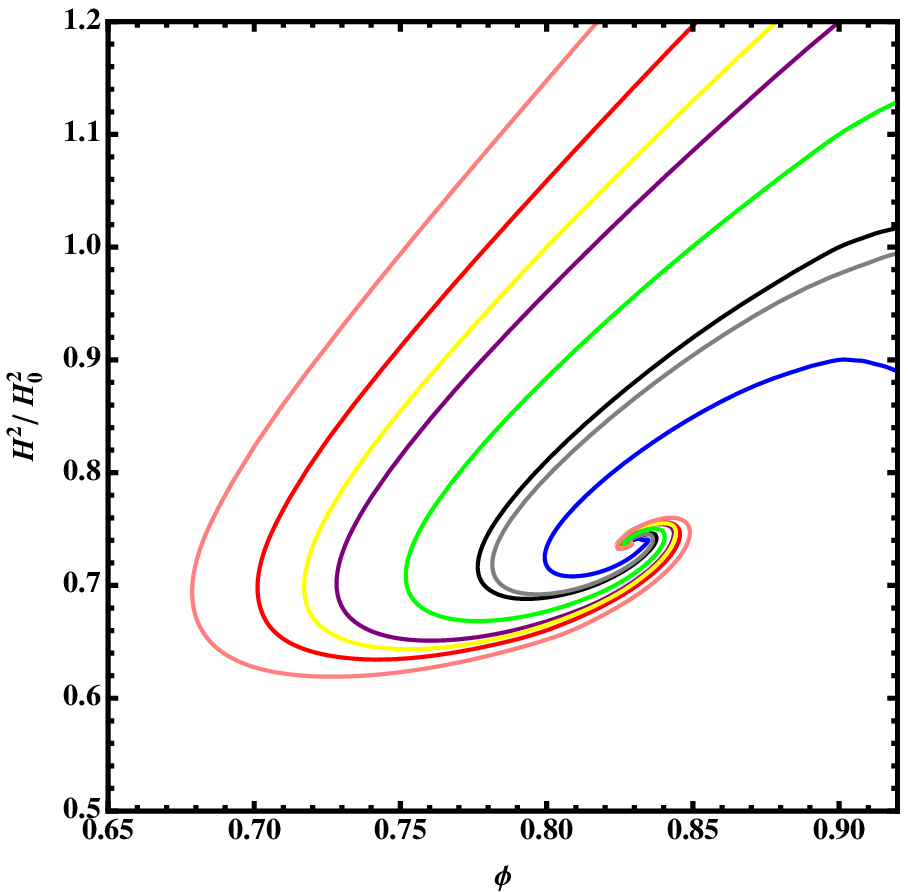}\label{fig9a}}\, \, \, \, \, \,
\subfloat[]{\includegraphics[width=0.40\textwidth]{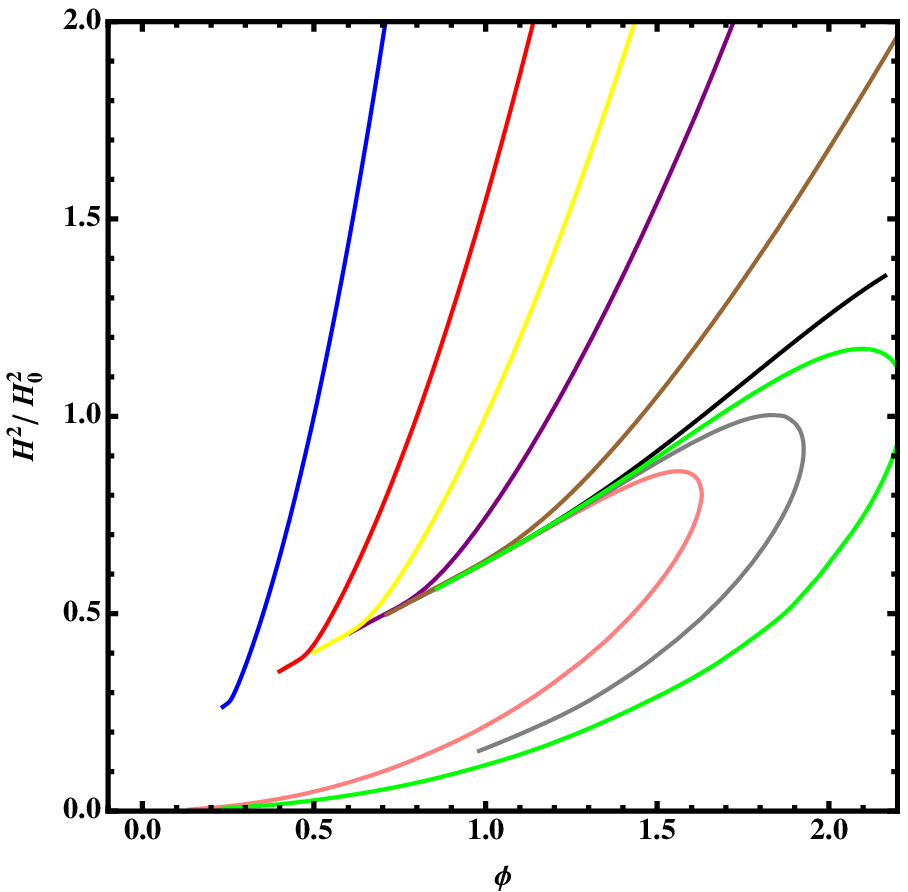}\label{fig9b}}
 \caption{\footnotesize{Phase space of the HS model (panel \ref{fig9a}), and the NO model (\ref{fig9b}). In both models, the potential $V_{-}(\phi)$ has been assumed. As suggested by the previous results, the HU model presents an asymptotically stable focus, whereas the NO model exhibits an asymptotically stable node.}}
\end{figure*}
This equation describes the phase space for a particular scalar potential, and provides useful information about the behavior and the stability of a particular model. At this step, let us analyze the phase space in vacuum for the $f(R)$ models (\ref{1.8a}) by assuming the negative branch of the scalar potential $V_{-}(\phi)$ given in (\ref{3.8}). Recall that for an illustrative propose, here $c_1=2$, $c_2=1$, and $a=0.1/H_0^2$ , $b=1$, $c=0.05/H_0^2$ are set for the HS and the NO model respectively. Then, it is straightforward to find that the only real critical points (\ref{3.9a}) are given by, 
\bea
\mbox{HS model:}\quad \phi_c=0.82\ ,\, H_c=0.74\ , \nn
\mbox{NO model:}\quad \phi_c=-5\ \times10^{-5}\ ,\, H_c\sim 0\ , \nn
\eea

Then, by analyzing  the dynamical system (\ref{3.9}), it is straightforward to find out that the critical point of the HS model corresponds to an asymptotically stable focus, as shown in Fig~\ref{fig9a}, whereas the critical point of the NO model is an asymptotic stable node, Fig.~\ref{fig9b}. Fig.~8 corresponds to the representation of the phase space of both models, where the black and gray curves correspond to the $\Lambda$CDM initial conditions and the phantom ones assumed above respectively. Moreover, other initial conditions are taken in order to illustrate the diagram. \\
As observed in Fig.~\ref{fig9a}, the HS model owns an stable de Sitter solution, which agrees with the results obtained in the previous section, whereas the NO model presents a stable point close to the origin, where the Hubble parameter becomes null. Nevertheless, note that the coefficients of the dynamical system of the NO model turns out complex for an specific value, so the Hubble parameter reach a limit, and most of the curves do not reach the origin but end at a finite non null value of the Hubble parameter, which leads asymptotically to a stable dS solution, that agrees with the results of the previous section. Hence, both models predict an asymptotic stable cosmological evolution. Nevertheless, in the next section, the other branch of the scalar potential is studied, which corresponds to another solution (in vacuum) of these $f(R)$ models, where the occurrence of a future singularity can not be avoided.


\section{Future singularities, Little Rip and Pseudo-Rip in viable $f(R)$ gravity}
\label{Singularities}

In recent years, the study of future singularities has become a major task since a lot of cosmological models, capable of describing a realistic cosmological evolution satisfying the observational constraints, lead to a phantom phase, where  $w<-1$, and may content some type of future singularity (see Refs.~\cite{phantom}). A  classification of future singularities was presented in Ref.~\cite{Nojiri:2005sx},
\begin{itemize}
\item Type I (``Big Rip''): For $t\rightarrow t_s$, $a\rightarrow
\infty$ and $\rho\rightarrow \infty$, $|p|\rightarrow \infty$.
\item Type II (``Sudden''): For $t\rightarrow t_s$, $a\rightarrow
a_s$ and $\rho\rightarrow \rho_s$, $|p|\rightarrow \infty$ (first pointed out in Ref.~\cite{barrow1} and studied in Ref.~\cite{Barrow:2004hk}).
\item Type III: For $t\rightarrow t_s$, $a\rightarrow a_s$
and $\rho\rightarrow \infty$, $|p|\rightarrow \infty$.
\item Type IV: For $t\rightarrow t_s$, $a\rightarrow a_s$ and
$\rho\rightarrow \rho_s$, $p \rightarrow p_s$
but higher derivatives of Hubble parameter diverge.
\end{itemize}

Future singularities have been widely studied, and some $f(R)$ models produce this divergent behavior (see Refs.~\cite{MyF(R)1,Nojiri:2008fk}). Nevertheless, the fact that the universe crosses the phantom barrier at $w=-1$ is not enough for the presence of a future singularity (see Ref.~\cite{Sahni:2002dx}). Moreover, future singularities may be cured using appropriate ingredients (see Ref.~\cite{LopezRevelles:2011uc}). In section \ref{Evolution}, we found out that the evolution of some particular viable models enters into a phantom epoch in the future, where the EoS parameter becomes less than -1, but no future singularity is detected since the Hubble parameter and its first derivative have a regular behavior. In section \ref{Scalar}, the study of the phase space confirmed that both models tend asymptotically to a stable dS solution. Nevertheless, note that the above viable models may present a singularity along the cosmological evolution, since  the first derivative of the scalar potential  contains a divergence at a finite value of $\phi$, precisely at the boundary of the scalar field defined by the scalar potentials (\ref{3.8}), which causes a sudden singularity, as pointed out in Ref.~\cite{Appleby:2009uf}.  As shown above, the scalar potential (\ref{3.8}) is split into two branches, where each one leads to different cosmological evolutions. Let us now analyze the positive branch $V_{+}(\phi)$ in both models. By assuming the numerical values for the free parameters of previous sections, it is straightforward to calculate the boundary of the scalar field, which yield $\phi_{bHS}=1$ and $\phi_{bNO}=3$, and where $dV(\phi)/d\phi$ diverges and a sudden singularity occurs. Fig.~9 corresponds to the phase diagram when $V_{-}$ is assumed, and all the curves end asymptotically at $\phi_{b}$ in both models . Hence, the evolution tend to a sudden singularity, such that branch of the scalar potential describes  another solution of the above viable $f(R)$ models, where a future singularity can not be avoided.  However, note that this analysis is done in absence of matter, while when is included, the cosmological evolution associated to the scalar potential $V_{-}(\phi)$ becomes the resulting solution in both $f(R)$ actions, leading to a stable cosmological evolution, as shown in the previous sections. \\
\begin{figure*}[h!]
	\centering
		\subfloat[]{\includegraphics[width=0.40\textwidth]{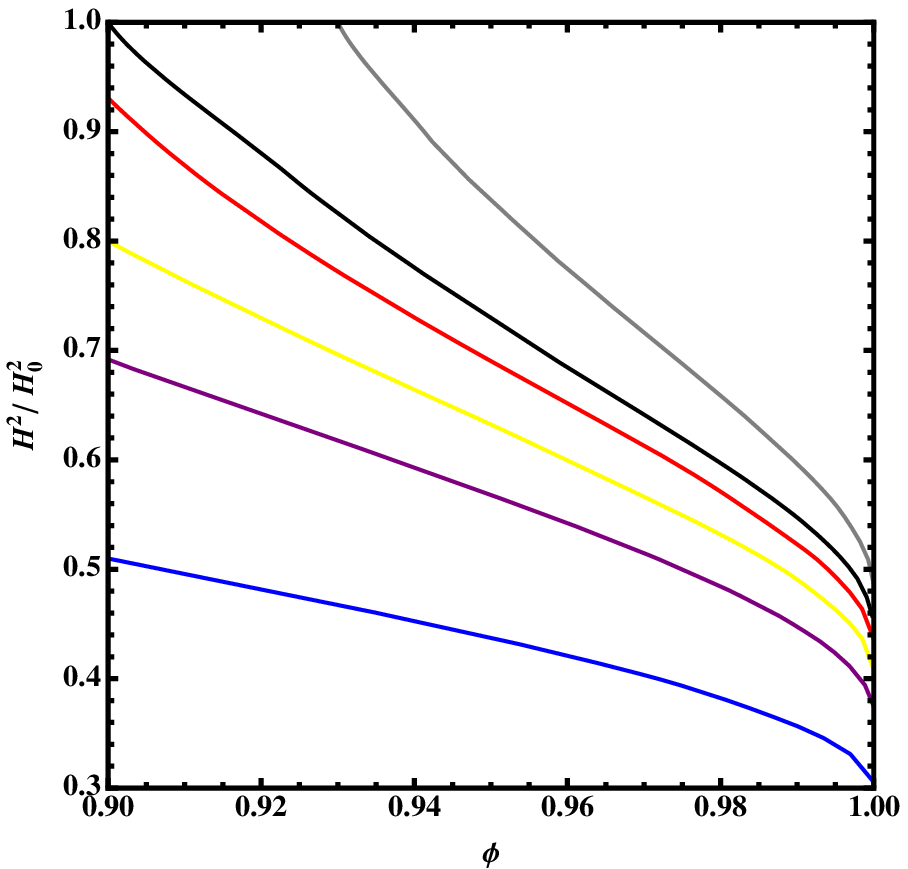}\label{fig10a}}\, \, \, \, \, \,
\subfloat[]{\includegraphics[width=0.40\textwidth]{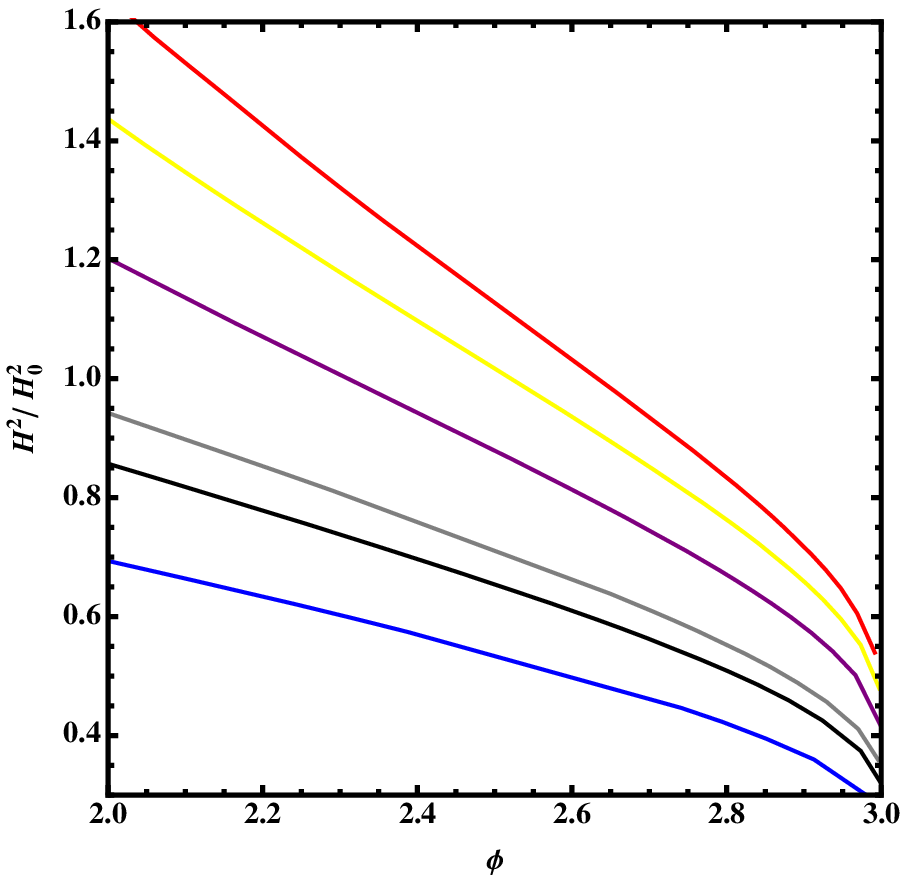}\label{fig10b}}
 \caption{\footnotesize{Phase space $H(\phi)$ for the HS model (panel \ref{fig10a}), where different initial conditions are considered. The right panel corresponds to the NO model. For both models, the potential $V_{+}(\phi)$ has been assumed. As observed, all the curves in both models reach the bounding of the scalar field, leading a sudden singularity.}}
\end{figure*}
 
Nevertheless, even if the presence of a future singularity is avoided,  other critical scenarios may occur, as the so-called {\it Little Rip} or the {\it Pseudo-Rip} scenario.  The {\it Little 
Rip} occurs when the Hubble parameter becomes an increasing function that diverges as time goes to infinity producing an inertial force that breaks bound systems, but where all physical magnitudes remain finite at finite time (see Ref.~\cite{LittleRip,Frampton:2011rh}). Moreover, in the {\it Pseudo-Rip}, the Hubble parameter remains finite, but the inertial force is capable of dissociating some bound structures along the cosmological evolution.  The effects of the expansion on local systems have been widely explored, which is still an open problem (for a review see Ref.~\cite{Carrera:2008pi}). Nevertheless, and except for the Einstein-Strauss model that imposes very restricted conditions,  the expansion of a FLRW background  seems to affect local systems (see Ref.~\cite{Mars:2008tq}). \\

The consequences of such scenarios would lead to the modification of the metrics that characterizes such bound systems, where the cosmological expansion would play an important role.  Here we are interested to study these scenarios in the viable $f(R)$ theories considered throughout this manuscript. Firstly, we should characterize, at least qualitatively, the strength of the inertial force induced by the cosmological expansion in order to be compared with those forces of a bound system. In this sense, the force that a body of mass $m$ may feel due to the cosmological expansion by the relative acceleration between two points separated a comoving distance $r$ can be expressed as,
\be
F_{cosm}=m\ r\ \left(\frac{\ddot{a}}{a}\right)=m\ r \left(H^2+\dot{H}\right)\ .
\label{LR1}
\ee
Then, according to the analysis of the models (\ref{1.8a}) of the previous sections, the Hubble parameter does not diverge at infinity since the evolution goes asymptotically to a stable de Sitter solution. Only the scalar potential $V_{+}$ induces a sudden singularity, but this is not the case in the presence of matter. Then,  a {\it Little Rip} can not occur, so let us consider the possibility of a {\it Pseudo Rip} scenario, by analyzing the evolution of the inertial force (\ref{LR1}) in comparison with the bounding force of a particular system. For an illustrative purpose, the Newtonian force of the coupling between the Sun and a body of mass $m$ is considered,
\be
F_{N}=G\frac{m M_{Sun}}{r^2}\ .
\label{LR2}
\ee
where $M_{Sun}=1.9\ 10^{30}\ kg$ is the mass of the Sun, and $r$ is the distance between the Sun and the body of mass $m$. In both cases $F_{i}=|\vec{F}_{i}|$ is assumed, since both forces have radial directions, but opposite sign. Hence, both quantities can be compared along the cosmological evolution,
\be
 \frac{F_{cosm}}{F_N}=\frac{r^3}{G\ M_{Sun}}H_0^2\left[h(z)^2-(1+z)h(z)\ h'(z)\right]\ ,
\label{LR3}
\ee
where recall that $H(z)=H_0\ h(z)$, where $H_0=100\, h \ km\ s^{-1}\ Mpc^{-1}$ with $h=0.71\pm 0.03$. For the case of the Earth-Sun system, $r=149.6\times 10^9\ m$,  the ratio (\ref{LR3}) yields $F_{cosm}/F_N=1.4\times 10^{-22}\left[h(0)^2-h(0)h'(0)\right]$ and by considering the same initial conditions of the previous sections, $h(0)=1$ and $a)$ $h'(0)=\frac{3}{2}\Omega_m^{(0)}$, and $b)$ $h'(0)=-0.1$, the ratio of both forces at $z=0$ is given by 
\be
\left( i\right)\  \frac{F_{cosm}}{F_N}\sim0.82\times 10^{-22}\ , \quad \left( ii\right)\  \frac{F_{cosm}}{F_N}\sim1.95\times 10^{-22}\ .
\label{LR5}
\ee
\begin{figure*}[h!]
	\centering
\subfloat[]{\includegraphics[width=0.40\textwidth]{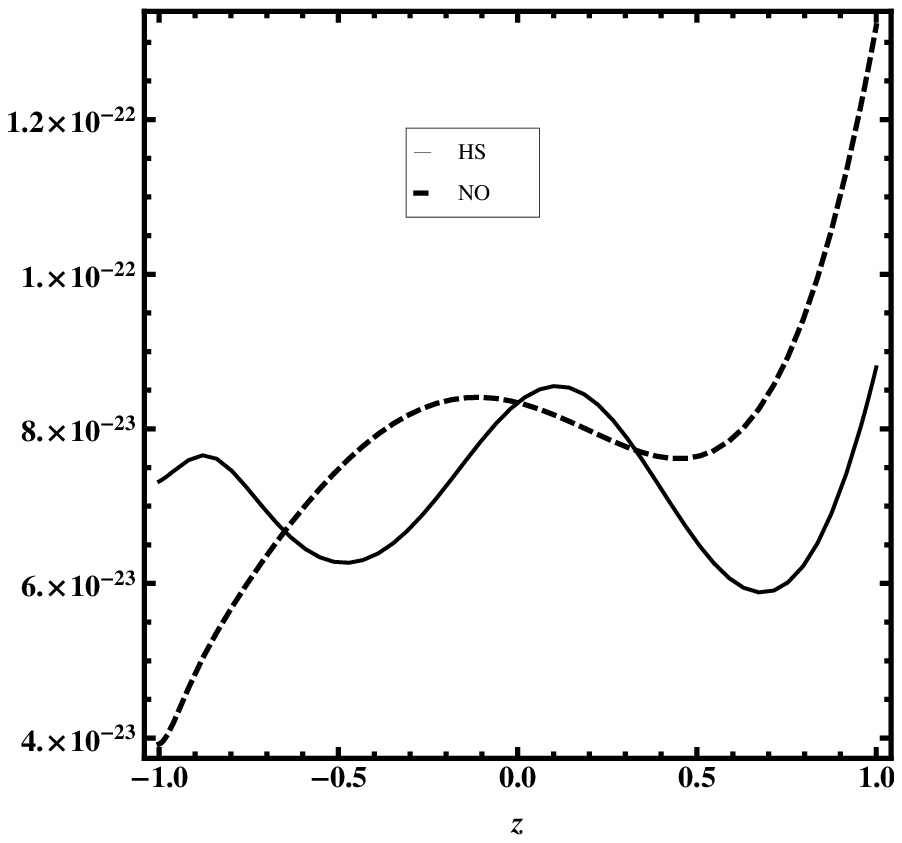}\label{fig11a}}\, \, \, \, \, \,
\subfloat[]{\includegraphics[width=0.40\textwidth]{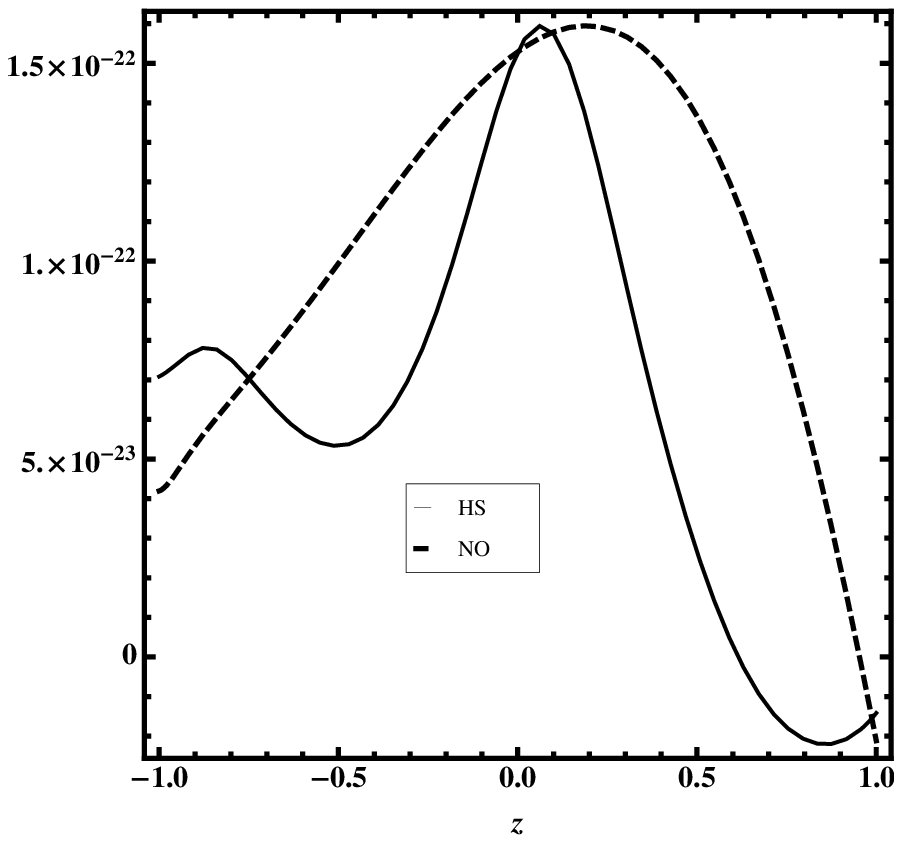}\label{fig11b}}
 \caption{\footnotesize{Evolution of the ratio $\frac{F_{cosm}}{F_N}$ for the Earth-Sun system for the HS and NO models. Here we assume $h(0)=1$ and: Panel $(a)$\ $h'(0)=\frac{3}{2}\Omega_m^{(0)}$, Panel$(b)$\  $h'(0)=-0.1$.}}
\end{figure*}
Hence, both values are too small to produce any dissociation of the Solar System, and in particular of the Earth orbit. In addition, the future cosmological evolution does not seem to affect significantly to the Earth-Sun system whatever the $f(R)$ model is assumed, and independently of the initial conditions imposed at $z=0$ as shown in Fig.~10. In the HS model the ratio (\ref{LR3}) behaves as an damped oscillator, and the amplitude is too small to affect the Newtonian force between the Earth and the Sun. In the  NO model, the expansion force reaches its maximum at positive redshifts in both figures, and decreases when the universe approximates to $z=-1$.\\ 

Hence, the viable $f(R)$ models considered here do not seem to affect  bound systems, and the {\it Pseudo-Rip} does not occur.
 
\section{Conclusions}


In the present paper, the so-called viable $f(R)$ models have been studied by analyzing two characteristic examples. We have found that the cosmological evolution of the effective EoS enters into a phantom stage in the present or near future for the free parameters assumed here. This analysis shows that both models can reproduce quite well the cosmological evolution, similarly to $\Lambda$CDM model, when the right conditions are assumed. As shown in section \ref{Evolution}, the cosmological evolution matches $\Lambda$CDM evolution when the standard initial conditions are set at $z=0$, whereas if the phantom transition is imposed to occur before $z=0$, the behavior of the cosmological parameters $\{q, \Omega_m,\Omega_F\}$  presents considerable deviations with respect to $\Lambda$CDM model. Nevertheless, the transition to the phantom stage occurs in both cases.\\
 
Moreover, such transition to a phantom epoch does not imply directly the occurrence of a future singularity, since the Hubble parameter and its first derivatives remain finite, even in the case that the transition to the phantom era had already occurred before $z=0$, according to the results obtained in section \ref{Evolution}. However, the choice of the initial conditions at $z=0$  affect naturally the past evolution, as shown in Figs.~3-5, where the evolution of the cosmological parameters  presents a large deviation from the $\Lambda$CDM model  at positive redshifts. \\

Nevertheless, this does not mean that these viable models are free of singularities, as shown by the analysis of the scalar-tensor counterpart, but the solution found in the presence of a dust fluid has a regular behavior in the range of redshifts explored here. In addition, the analysis of the equivalence scalar-tensor theory reveals that these $f(R)$ models  are split into two branches through the scalar potential, and  a boundary  on the scalar field results. The analysis of the phase space in vacuum reveals that this kind of models  owns very different behaviors, whereas one of the branches of the scalar potential presents an asymptotically stable evolution as illustrated in Fig.~8, resulting in a stable dS solution in both models, the other branch ends asymptotically in the boundary of the scalar field, where a sudden singularity occurs \cite{Appleby:2009uf}. However, in the presence of dust matter and with the numerical values assumed here, the stable branch becomes the resulting solution, encouraging the idea that future singularities can be avoided, and a asymptotic stable behavior can be achieved in viable $f(R)$ gravities.\\

In the last section, we have also discussed the possibility of the occurrence of a {\it Little Rip} and a  {\it Pseudo-Rip}, which refer to  stages of the universe evolution that may induce a disassociation of some bound systems due to the strength of the expansion. This possibility has been qualitatively explored by comparing the strength felt by a body with respect to another point of the universe, and the newtonian force of a typical planet system. For the Earth-Sun system, the expansion force is too small to produce any significant effect at a local system, so none of the above scenarios occur. \\ 

Hence, in the present paper, we have realized a deep analysis on viable $f(R)$ gravities, showing that the models considered here are capable of reproducing the late-time acceleration, entering in a phantom phase in the present or near future, a fact that may distinguish from other dark energy models. Furthermore, the analysis of the phase space in the scalar-tensor counterpart  reveals a regular and irregular behavior of these models, where the imposition of the regular solution provides a way to obtain a viable $f(R)$ gravity with an asymptotic stable evolution.

\ack

I would like to thank Sergei Odintsov, Alvaro de la Cruz-Dombriz and Ra\"ul Vera for useful discussions on the topic. I acknowledge support from a postdoctoral contract from the University of the Basque Country (UPV/EHU) under the program ``Specialization of research staff'', and support from the research project FIS2010-15640, and also by the Basque Government through the special research action KATEA and UPV/EHU under program UFI 11/55.

  \end{document}